\documentclass{elsartL}
\usepackage{graphicx}
\usepackage{amssymb}
\usepackage{amsmath}
\usepackage{physics}
\usepackage{mathrsfs}
\usepackage{mathtools}
\usepackage{appendix}
\newcommand{\avg}[1]{\left< #1 \right>} 
\newcommand\at[2]{\left.#1\right|_{#2}}
\begin{document}

\begin{frontmatter}
\title{Comparison of results for the electromagnetic form factors of the proton at low $Q^2$}
\author{Evangelos Matsinos}

\begin{abstract}The goal in this work is the comparison of five parameterisations of the Sachs form factors of the proton $G^p_E$ (electric) and $G^p_M$ (magnetic) at low $4$-momentum transfer $Q^2$. It will be shown that a 
simple model, based on two dipoles which admit as parameters the rms electric charge radius $r_{E;p}$ and the rms magnetic radius $r_{M;p}$ of the proton, suffices for the purposes of the phase-shift analyses (PSAs) of the 
low-energy pion-nucleon ($\pi N$) data. The replacement of the electromagnetic form factors, currently used in the ETH model of the $\pi N$ interaction, by the parameterisation of this work will enable the removal from the 
PSAs of this research programme of the largest part of the dependence on extraneous sources.\\
\noindent {\it PACS 2010:} 13.40.-f, 13.40.Gp, 14.20.Dh
%
\end{abstract}
\end{frontmatter}

\section{\label{sec:Introduction}Introduction}

To determine the scattering amplitude of electromagnetic (EM) processes, pertaining to the elastic scattering of charged particles on proton targets, an expression for the EM transition current of the proton is required. The 
general form of the proton vertex, fulfilling Lorentz invariance and charge conjugation, can be found in Chapter 8.8.2 of Ref.~\cite{Aitchison2013}, p.~259:
\begin{equation} \label{eq:EQ0001}
\bra{p^\prime, s^\prime} J^\mu_{\rm EM} \ket{p, s} = e \, \bar{u} (p^\prime, s^\prime) \, \left( F^p_1 \gamma^\mu + i \frac{\kappa_p F^p_2}{2 m_p} \sigma^{\mu \nu} q_\nu \right) \, u(p,s) \, \, \, ,
\end{equation}
where
\begin{itemize}
\item $e$ is the electric charge of the proton;
\item $m_p$ is the proton mass;
\item $p$ and $s$ stand for the $4$-momentum and spin of the initial-state proton;
\item $p^\prime$ and $s^\prime$ stand for the $4$-momentum and spin of the final-state proton;
\item $u (p, s)$ is the Dirac spinor associated with the plane-wave of a proton with $4$-momentum $p$ and spin $s$;
\item the quantities $\gamma^{\mu}$ ($\mu = 0, 1, 2, 3$) are the standard Dirac $4 \times 4$ matrices, satisfying the relation $\{ \gamma^{\mu}, \gamma^{\nu} \} = 2 g^{\mu \nu} I_4$, $g^{\mu \nu}$ being the Minkowski metric, 
with signature `$+ \, - \, - \, -$';
\item the matrices $\sigma^{\mu \nu}$ are defined by the relation: $\sigma^{\mu \nu} = \frac{i}{2} [\gamma^\mu , \gamma^\nu]$; and
\item $q = p^\prime - p$ represents the $4$-momentum transfer (i.e., the $4$-momentum of the EM current). The standard Mandelstam variable $t$ is defined as: $t \coloneqq q^\mu q_\mu = q^2$. As $t \leq 0$ in the 
physical region for elastic scattering, widely used in Particle Physics is the $4$-momentum transfer in the form $Q^2 \coloneqq -t \geq 0$.
\end{itemize}
The quantity $\kappa_p \coloneqq \mu_p - 1$ is known as `anomalous magnetic moment' of the proton~\footnote{The word `anomalous' indicates that $\kappa_p$ is the magnetic moment in excess of $1$; for a structureless proton, 
$\kappa_p$ vanishes.}, where $\mu_p$ is the numerical value of the magnetic moment of the proton, when expressed in units of the nuclear magneton $\mu_N \coloneqq e \hbar / (2 m_p)$. Recommended by the Particle Data Group 
(PDG) \cite{pdg2020} is $\mu_p=2.79284734462(82)$, taken from Ref.~\cite{Schneider17}, a value which will be used in the numerical results of this work.

In Eq.~(\ref{eq:EQ0001}), the Dirac $F^p_1$ and Pauli $F^p_2$ form factors are $t$-dependent functions. According to an older convention (which is also followed in Ref.~\cite{Aitchison2013}), these two form factors were taken 
to satisfy the normalisation conditions: $F^p_1(0)=F^p_2(0)=1$. Another convention has gained popularity at more recent times: the constant $\kappa_p$ is now usually absorbed in $F^p_2 (t)$, thus yielding the normalisation 
condition: $F^p_2(0) = \kappa_p$. As this redefinition of $F^p_2 (t)$ (i.e., $\kappa_p F^p_2 (t) \to F^p_2 (t)$) somewhat simplifies the expressions for scattering, the recent convention will be adopted in this paper.

In the Born approximation (one-photon exchange), the differential cross section, describing electron-proton ($e p$) elastic scattering in the laboratory frame of reference, was put into the form
\begin{equation} \label{eq:EQ0002}
\frac{d \sigma}{d \Omega} = \left( \frac{d \sigma}{d \Omega} \right)_{\rm ns} \left( (F^p_1)^2 + \tau (F^p_2)^2 + 2 \tau \left( F^p_1 + F^p_2 \right)^2 \tan^2 (\theta/2) \right)
\end{equation}
by Rosenbluth \cite{Rosenbluth1950} (also see Eq.~(8.207) in Ref.~\cite{Aitchison2013}, p.~259), where $(d \sigma / d \Omega)_{\rm ns}$ represents the so-called `no-structure' differential cross section, frequently referred 
to as `Mott cross section', see Eq.~(8.49) in Ref.~\cite{Aitchison2013}, p.~229. (Not included in Eq.~(\ref{eq:EQ0002}) are the proton-recoil effects.) The quantity $\tau$ in Eq.~(\ref{eq:EQ0002}) is defined as the ratio 
$Q^2 / (4 m^2_p)$. Finally, $\theta$ is the scattering angle of the projectile (electron).

The two form factors of the neutron $F^n_1$ and $F^n_2$ are defined similarly, and (of course) are also $t$-dependent. The Dirac and Pauli form factors of the nucleons were first expressed in terms of the so-called EM Sachs 
(electric $G^N_E$ and magnetic $G^N_M$) form factors in the early 1960s \cite{Hand1963}:
\begin{equation} \label{eq:EQ0003}
F^N_1 = \frac{G^N_E + \tau G^N_M}{1 + \tau}
\end{equation}
and
\begin{equation} \label{eq:EQ0004}
F^N_2 = \frac{G^N_M - G^N_E}{1 + \tau} \, \, \, ,
\end{equation}
where $N=p$ or $n$. From Eqs.~(\ref{eq:EQ0003},\ref{eq:EQ0004}), one obtains
\begin{equation} \label{eq:EQ0005}
G^N_E = F^N_1 - \tau F^N_2
\end{equation}
and
\begin{equation} \label{eq:EQ0006}
G^N_M = F^N_1 + F^N_2 \, \, \, .
\end{equation}
Evidently, the two Sachs form factors of the proton satisfy the normalisation conditions: $G^p_E(0)=1$ and $G^p_M(0)=\mu_p$. For the corresponding quantities of the neutron: $G^n_E(0)=0$ (the neutron has no net electric charge) 
and $G^n_M(0)=\mu_n$ (equal to $-1.91304273(45)$ \cite{Mohr2016}, according to the PDG recommendation \cite{pdg2020}).

\begin{figure}
\begin{center}
\includegraphics [width=15.5cm] {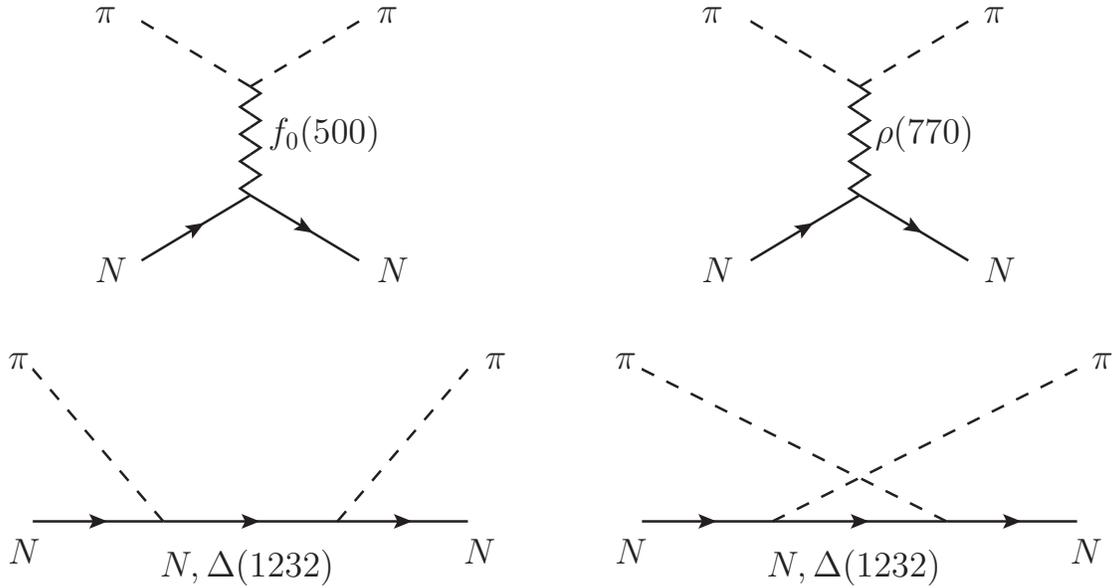}
\caption{\label{fig:FeynmanGraphsETHZ}The main Feynman graphs of the ETH model: scalar-isoscalar $I^G\,(J^{PC}) = 0^+\,(0^{++})$ and vector-isovector $I^G\,(J^{PC}) = 1^+\,(1^{--})$ $t$-channel graphs (upper part), and $N$ 
and $\Delta (1232)$ $s$- and $u$-channel graphs (lower part). Not shown in this figure, but also analytically included in the model, are the small contributions from all other scalar-isoscalar and vector-isovector mesons 
\cite{Matsinos2020a} with rest masses below $2$ GeV, as well as from all well-established (four-star) $s$ and $p$ higher baryon resonances (HBRs) \cite{Matsinos2020b} in the same mass range.}
\vspace{0.35cm}
\end{center}
\end{figure}

Relevant in the context of a pion-nucleon ($\pi N$) interaction model \cite{Goudsmit1994,Matsinos2006,Matsinos2014,Matsinos2017} (ETH model, henceforth), which accounts for the strong-interaction (hadronic) part of the $s$- 
and $p$-wave scattering amplitudes on the basis of $s$-, $u$-, and $t$-channel Feynman diagrams (see Fig.~\ref{fig:FeynmanGraphsETHZ}), are the proton (and pion) form factors at pion laboratory kinetic energy $T \leq 100$ MeV. 
The restriction on $T$ (necessitated by several reasons, e.g., see Section 1 of Ref.~\cite{Matsinos2017}) imposes an upper limit on the $Q^2$ values involved in the phase-shift analyses (PSAs) of the ETH model. The 
$Q^2_{\rm max}$ value is attained at $T = 100$ MeV and backward scattering: $Q^2_{\rm max} \approx 98941.3$ MeV$^2 \approx 0.1$ GeV$^2$. On the other hand, the available experimental data, used as input in determinations of 
the form factors $G^N_E$ and $G^N_M$, span a $Q^2$ domain exceeding $10$ GeV$^2$; therefore, of interest to the ETH model is very small part of the $Q^2$ domain within which $e p$ experimental data are available.

\clearpage

In this work, the masses and the $4$-momenta will be expressed in energy units. As a result, some of the expressions herein, involving the $4$-momentum of the EM current, will contain the conversion constant $\hbar c$, rather 
than the reduced Planck constant $\hbar$.

\section{\label{sec:Parameterisations}Parameterisation schemes at low $Q^2$}

The $t$-dependence of the Sachs form factors cannot be derived from first principles \cite{Hill2010}. Although this is not particularly problematic as far as the description of the experimental data is concerned, it becomes 
problematic when one attempts to extract from the fitted parameters (of an assumed parameterisation) physical quantities pertaining to properties of the nucleons (see Section 5.3 of Ref.~\cite{Sick2018}). For the sake of 
example, regarding the extraction of the rms electric charge radius of the proton from parameterisations of $G^p_E$, criticism has appeared in Refs.~\cite{Sick2018,Sick2014}. As Sick and Trautmann pointed out \cite{Sick2014}, 
while such parameterisations are ``valid representations of the data in the $q$-region where they have been measured, they are not suitable for an extrapolation to $q = 0$ where the proton rms-radii are extracted.'' Section 
3 of Ref.~\cite{Sick2018} provides a detailed account of the complexity in this issue, whereas Section 5.3 therein starts with the remark: ``Due to the complications mentioned in Section 3, most authors analyzing the electron 
scattering data employ parameterizations in $q$-space only to get the $q = 0$ slope, without ever worrying what these parameterizations would imply in $r$-space.''

The conundrum lies in the fact that the functional behaviour of $G^p_E (q)$ (a momentum-space attribute) and of the charge density $\rho (r)$ (a configuration-space attribute), representing the particle proton, are related via 
a Fourier transformation. Regarding the former, one can write \cite{Sick2018}:
\begin{equation} \label{eq:EQ0007}
G^p_E (q) = \frac{4 \pi \hbar c}{q} \int_{0}^\infty \rho(r) \sin \left( \frac{q r}{\hbar c} \right) r dr \, \, \, ,
\end{equation}
whereas for the latter, the inverse transformation yields:
\begin{equation} \label{eq:EQ0008}
\rho (r) = \frac{1}{2 \pi^2 r (\hbar c)^2} \int_{0}^\infty G^p_E (q) \sin \left( \frac{q r}{\hbar c} \right) q dq \, \, \, .
\end{equation}
Although both expressions are valid when the recoil velocity of the proton is small ($\beta \ll 1$), relativistic corrections are available, see Section 3 of Ref.~\cite{Sick2018}. In several parameterisations of $G^p_E$, the 
large-$r$ tail of $\rho$, obtained from Eq.~(\ref{eq:EQ0008}), clashes with our understanding of the proton \cite{Sick2018}.

In fact, starting `from the other end' appears to be promising \cite{Sick2014}: one could assume $\rho (r)$ distributions, which make physical sense, and obtain the corresponding parameterisations of $G^p_E$ via Eq.~(\ref{eq:EQ0007}). 
The recoil effects could then be treated as described in Section 3 of Ref.~\cite{Sick2018}.

My main interest in this work concerns the parameterisation of the Sachs form factors of the proton at low $Q^2$ for the purposes of PSAs conducted with the ETH model. At present, I have no intention to make contributions 
to the problem of the determination of the rms radii of the proton. Therefore, I will next list some of the available parameterisations (in chronological order of appearance), without entering the subject of the large-$r$ 
behaviour of $\rho$. The parameterisations of Sections \ref{sec:StandardDipole} and \ref{sec:ThisWork} follow dipole models, known to provide a good description of the nucleon form factors at low $Q^2$ values, and also (at 
least in case of $G^n_M$) at moderate ones, e.g., see Ref.~\cite{Lachniet2009}.

\subsection{\label{sec:StandardDipole}The `standard dipole' parameterisation}

Being the product of systematic experimentation by Hofstadter~\footnote{For ``his pioneering studies of electron scattering in atomic nuclei and for his thereby achieved discoveries concerning the structure of the nucleons,'' 
Hofstadter was awarded the Nobel Prize in Physics 1961.} and collaborators at Stanford University, the `standard dipole' form \cite{Hand1963}
\begin{equation} \label{eq:EQ0009}
f_D (t) = \left( 1 - \frac{t}{\Lambda^2} \right)^{-2} \, \, \, ,
\end{equation}
with $\Lambda^2 = 0.71$ GeV$^2$, enabled the routine parameterisation of the Sachs form factors of the nucleon $G^N_E$ and $G^N_M$ up to the 1980s.

One notable application of the `standard dipole' forms, relevant to the development of the ETH model in the early 1990s, was in the programme of the NORDITA team, which led to the extraction of the EM corrections \cite{Tromborg1976,Tromborg1977,Tromborg1978}, 
suitable for the analysis of the $\pi^\pm p$ scattering data; these corrections extend up to a momentum (in the centre-of-momentum (CM) frame of reference) equal to three times the charged-pion mass, equivalent to $T \approx 531$ 
MeV. Regarding their formulae (one-photon-exchange contribution to the EM scattering amplitude), one needs to pay some attention as Tromborg, Waldestr{\o}m, and {\O}verb{\o} abide by the NORDITA definition of the Pauli form 
factor, which is different to everyone else's: $(F^N_2)^{\rm NORDITA}= F^N_2 / (2 m_p)$, where $F^N_2$ is the Pauli form factor of this work~\footnote{For the `nucleon mass' $m_N$, the NORDITA team assumed in their works: 
$m_N = m_p$; others favour: $m_N=(m_p+m_n)/2$.}. The NORDITA parameterisation of the Sachs form factors of the nucleon followed the scheme \cite{Tromborg1974}:
\begin{equation} \label{eq:EQ0010}
G^p_E (t) = \frac{G^p_M (t)}{\mu_p} = \frac{G^n_M (t)}{\mu_n} = f_D (t) \, \, \, , G^n_E (t) = 0 \, \, \, .
\end{equation}
For the pion form factor $F^\pi$, NORDITA used \cite{Tromborg1974}: $F^\pi (t) = F^p_1 (t) - F^n_1 (t)$.

\subsection{\label{sec:A1}A parameterisation from the A1 Collaboration}

In 2010, the A1 Collaboration published the results of an analysis of differential cross-section and polarisation $e p$ measurements acquired at the Mainz Microtron (MAMI) \cite{Bernauer2010}. Their more detailed 2014 paper 
\cite{Bernauer2014} included important information on their parameterisations of the Sachs form factors of the proton $G^p_E$ and $G^p_M$. The results of their fits had appeared already in 2010 as material supplementing their 
first paper. Chosen herein are the fitted values of $G^p_E (t)$ and $G^p_M (t)$, obtained on the basis of spline fits to the World data; according to Table IV of Ref.~\cite{Bernauer2014}, their spline fits yield the best 
description of the input database. Some criticism about the results of Ref.~\cite{Bernauer2010} is expressed in Section 4 of Ref.~\cite{Sick2018}.

\subsection{\label{sec:VAMZ}The VAMZ parameterisation}

In 2007, Arrington, Melnitchouk, and Tjon \cite{Arrington2007} extracted $G^p_E (t)$ and $G^p_M (t)$ from constrained fits to the available cross-section and polarisation $e p$ measurements, including corrections accounting 
for two-photon-exchange effects \cite{Blunden2003}. The functions $G^p_E$ and $G^p_M$ were parameterised using a Pad{\'e} approximant of order $[3/5]$, namely
\begin{equation} \label{eq:EQ0011}
G^p_E (t), G^p_M (t) / \mu_p = \frac{1 + \sum_{i=1}^3 a_i \tau^i}{1 + \sum_{i=1}^{5} b_i \tau^i} \, \, \, .
\end{equation}
The fitted values of the parameters $a_i$ and $b_i$ for the two form factors were presented in tabulated form (see Table I of Ref.~\cite{Arrington2007}).

A few years later, Venkat, Arrington, Miller, and Zhan \cite{Venkat2011} used again the Pad{\'e} parameterisation of Eq.~(\ref{eq:EQ0011}) - with renamed parameters ($q_i$ for $G^p_E$ and $p_i$ for $G^p_M / \mu_p$) - as well 
as an improved theoretical background. The PSAs of the low-energy $\pi N$ data with the ETH model since 2015 have been based on the Sachs form factors of the proton obtained in Ref.~\cite{Venkat2011}. For the sake of brevity, 
this parameterisation will be named `VAMZ' henceforth.

\subsection{\label{sec:YAHL}The YAHL parameterisation}

Arrington's team employed another parameterisation of the Sachs form factors of the nucleon $G^N_E$ and $G^N_M$ in 2018 \cite{Ye2018}, this time in terms of the so-called $z$-expansion; for the sake of brevity, this 
parameterisation will be named `YAHL' henceforth. Their model rests upon high-degree polynomials (twelve-degree polynomials are used for the proton form factors, ten-degree polynomials for the neutron form factors) in the 
variable $z$, representing a conformal mapping of $t$ onto the unit circle:
\begin{equation} \label{eq:EQ0012}
z = \frac{\sqrt{t_{\rm cut} - t} - \sqrt{t_{\rm cut} - t_0}}{\sqrt{t_{\rm cut} - t} + \sqrt{t_{\rm cut} - t_0}} \, \, \, ,
\end{equation}
where $t_{\rm cut} = 4 m^2_c$ (two-pion cut), whereas the free parameter $t_0$ (the root of $z(t)=0$) was globally fixed in Ref.~\cite{Ye2018} to $-0.7$ GeV$^2$. The authors tabulated their results in the document `Explanation 
of Supplementary Material', also taking the trouble to detail the input to their optimisation scheme and make it available to others as supplementary material to their paper. Albeit straightforward, the authors also provided 
the code relating to the implementation of their results in two standard computer languages.

\subsection{\label{sec:ThisWork}The parameterisation of this work}

To model the $t$-dependence of the Sachs form factors of the proton $G^p_E$ and $G^p_M$, two dipoles will be introduced in this work, featuring two parameters, namely the rms electric charge radius $r_{E;p}$ of the proton and 
its rms magnetic radius $r_{M;p}$. The two radii are defined on the basis of the Sachs form factors of the proton according to the relations:
\begin{equation} \label{eq:EQ0013}
r^2_{E,M;p} \equiv \avg{r^2_{E,M}}_p \coloneqq \frac{6 (\hbar c)^2}{G^p_{E,M} (0)} \at{\frac{d G^p_{E,M} (t)}{d t}}{t=0} \, \, \, .
\end{equation}
As
\begin{equation} \label{eq:EQ0014}
G^p_{E,M} (t) = G^p_{E,M} (0) \left( 1 - \frac{t}{\Lambda^2_{E,M;p}} \right)^{-2} \, \, \, ,
\end{equation}
one obtains
\begin{equation} \label{eq:EQ0015}
\Lambda_{E,M;p} = \frac{2 \sqrt{3} \hbar c}{r_{E,M;p}} \, \, \, .
\end{equation}

For the sake of completeness, regarding the pion form factor $F^\pi$ (which is also required in the EM part of the $\pi N$ interaction), a monopole is being used in the ETH model since 2015. (In fact, it does not matter much 
whether the pion form factor is parameterised at low $Q^2$ according to the dipole or to the monopole model.)
\begin{equation} \label{eq:EQ0016}
F^\pi (t) = \left( 1 - \frac{t}{\Lambda^2_{E;\pi}} \right)^{-1} \, \, \, ,
\end{equation}
where
\begin{equation} \label{eq:EQ0017}
\Lambda_{E;\pi} = \frac{\sqrt{6} \hbar c}{r_{E;\pi}} \, \, \, ,
\end{equation}
which is the equivalent of Eqs.~(\ref{eq:EQ0015}) in case of a monopole model.

\section{\label{sec:Results}Results}

\begin{figure}
\begin{center}
\includegraphics [width=15.5cm] {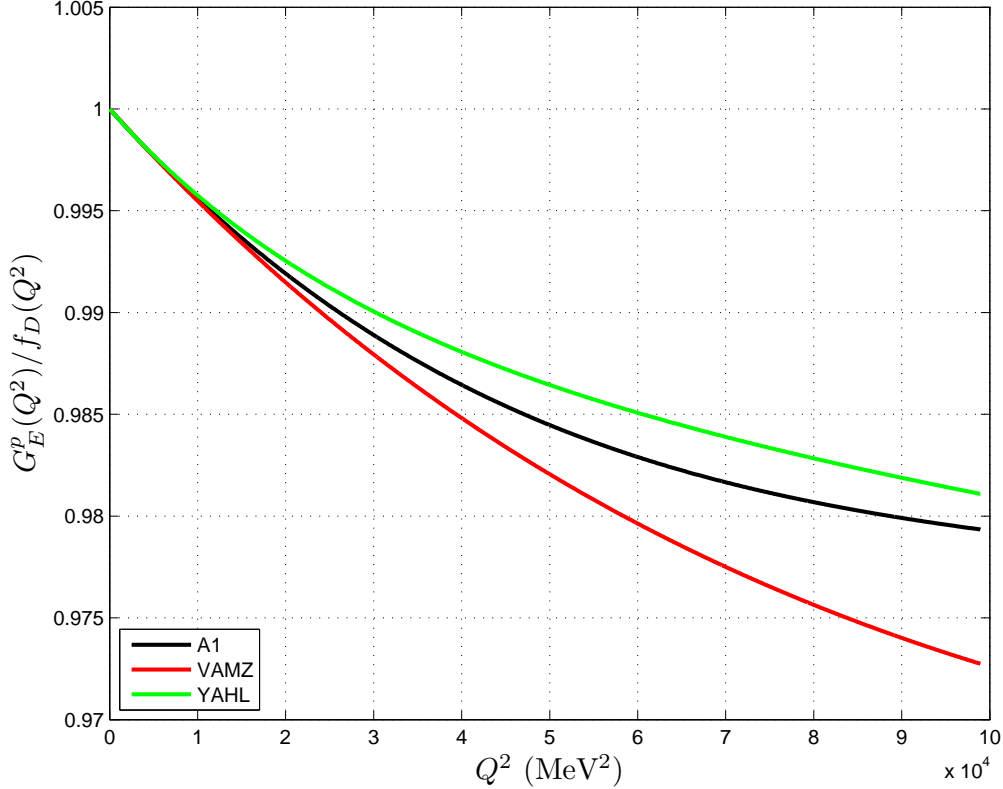}
\caption{\label{fig:GpE}The dependence of the Sachs form factor of the proton $G^p_E$ on the square of the $4$-momentum transfer $Q^2 \coloneqq -t$, where $t$ is the standard Mandelstam variable. The $Q^2$ range corresponds 
to the domain of interest in the context of the ETH model of the $\pi N$ interaction, corresponding to pion laboratory kinetic energy $T \leq 100$ MeV. To reduce the range of variation of $G^p_E$, ratios are shown of the 
parameterisations of Sections \ref{sec:A1}-\ref{sec:YAHL} to the values obtained with the `standard dipole' of Section \ref{sec:StandardDipole}.}
\vspace{0.35cm}
\end{center}
\end{figure}

\begin{figure}
\begin{center}
\includegraphics [width=15.5cm] {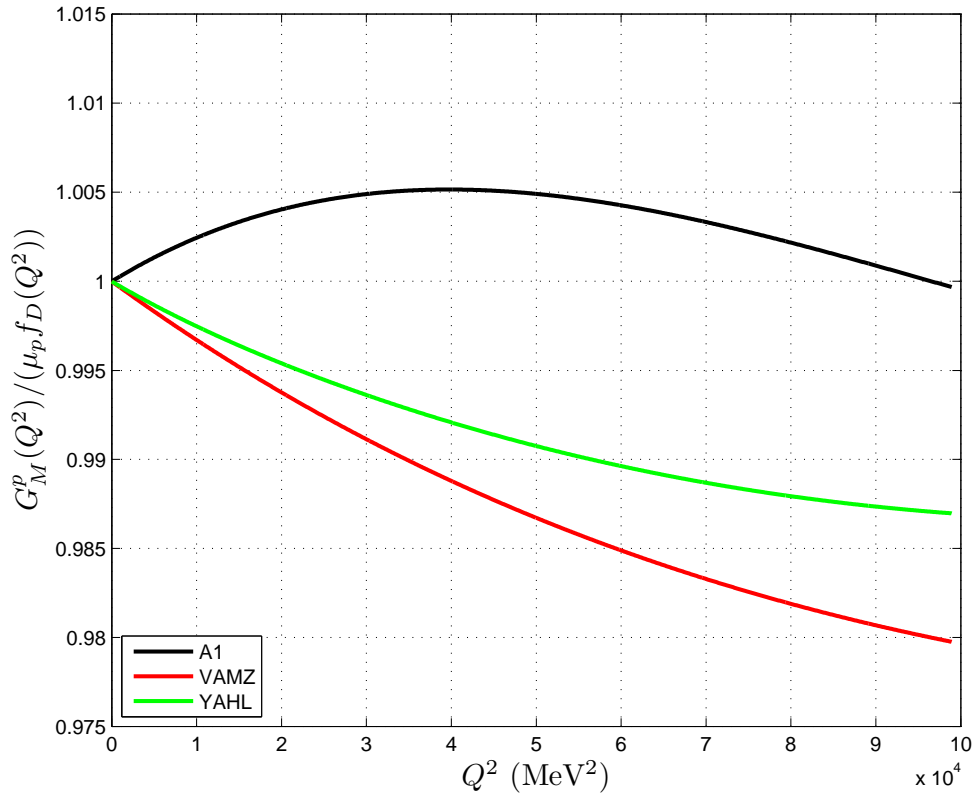}
\caption{\label{fig:GpM}The equivalent of Fig.~\ref{fig:GpE} for the Sachs form factor of the proton $G^p_M$.}
\vspace{0.35cm}
\end{center}
\end{figure}

The parameterisations of the proton form factors of Sections \ref{sec:StandardDipole}-\ref{sec:YAHL} are `fixed in time', in that they have been obtained on the basis of certain educated guesses for their parameters (`standard 
dipole') or from fits to available data (A1, VAMZ, YAHL), where several other physical constants had been imported from extraneous sources. The results for the Sachs form factors of the proton, obtained from these 
parameterisations, are compared in Figs.~\ref{fig:GpE} and \ref{fig:GpM}.

The maximal symmetric mean absolute differences~\footnote{The symmetric mean absolute difference between two values $v_1$ and $v_1$ is defined equal to $2 \lvert v_1 - v_2 \rvert / (\lvert v_1 \rvert + \lvert v_2 \rvert)$.} 
in $G^p_E$ among these results range (in the $Q^2$ domain of the plot) between $0.23~\%$ (A1-YAHL) and $2.76~\%$ (`standard dipole'-VAMZ). Regarding $G^p_M$, the differences range between $0.51~\%$ (`standard dipole'-A1) and 
$2.05~\%$ (A1-VAMZ). In comparison with the other three solutions, the `standard dipole' systematically overestimates $G^p_E$ in the $Q^2$ domain of Fig.~\ref{fig:GpE}. A similar effect is observed in Fig.~\ref{fig:GpM} for 
the two solutions originating from Arrington's team. Regarding the low-$Q^2$ behaviour of the A1 solution for $G^p_M$, one notices that it exceeds all other solutions, including the one corresponding to the `standard dipole' 
up to about $Q^2_{\rm max}$. This behaviour of the A1 solution for $G^p_M$ is reflected in a sizeable difference (to all other solutions) in $r_{M;p}$; as inspection of Table \ref{tab:Comparison} (which will be introduced 
shortly) reveals, the $r_{M;p}$ value in the A1 solution is about $10~\%$ lower than the values associated with the parameterisations of Sections \ref{sec:StandardDipole}, \ref{sec:VAMZ}, and \ref{sec:YAHL}.

To be able to include in the comparison the proton form factors from the parameterisation of this work, one would first need to fix the model parameters $r_{E;p}$ and $r_{M;p}$. Before that, however, one test might be helpful: 
one could first assess the differences between (on the one hand) the parameterisations of Sections \ref{sec:StandardDipole}-\ref{sec:YAHL} and (on the other hand) the parameterisation of this work after fixing the model parameters 
$r_{E;p}$ and $r_{M;p}$ to the appropriate values corresponding to each of the former four parameterisation schemes. The $r_{E;p}$ and $r_{M;p}$ values which have been used in this comparison, along with the results for the 
maximal symmetric mean absolute difference $d_{\rm max}$ in the $Q^2$ domain of this work, are given in Table \ref{tab:Comparison}. Regarding the A1, VAMZ, and YAHL parameterisations, the $r_{E;p}$ and $r_{M;p}$ results can be 
estimated from the parameter values found in Refs.~\cite{Bernauer2010,Venkat2011,Ye2018}. For the sake of example, $r_{E;p} = \hbar c \sqrt{6 (q_2 - q_6)} / (2 m_p)$ in the VAMZ parameterisation, and a similar expression holds 
for $r_{M;p}$ after the replacement of the parameters: $q_{2,6} \to p_{2,6}$.

\begin{table}
{\bf \caption{\label{tab:Comparison}}}Comparison of the results obtained with the parameterisation of the Sachs form factors of the proton $G^p_E$ and $G^p_M$ of this work (see Section \ref{sec:ThisWork}) with the form factors 
detailed in Sections \ref{sec:StandardDipole}-\ref{sec:YAHL}. The maximal symmetric mean absolute differences $d_{\rm max}$ correspond to the $Q^2$ domain of this work, namely up to about $0.1$ GeV$^2$. In this comparison, the 
model parameters $r_{E;p}$ and $r_{M;p}$ have been fixed (separately for the purposes of each comparison) to the quoted values per case, which correspond to the results of the parameterisations of Sections \ref{sec:StandardDipole}-\ref{sec:YAHL}. 
Before taking the $r_{E,M;p}$ values of this table too seriously, the reader should bear in mind the criticism expressed in Ref.~\cite{Sick2018}, as outlined at the beginning of Section \ref{sec:Parameterisations} of 
this work.
\vspace{0.3cm}
\begin{center}
\begin{tabular}{|c|c|c|c|c|}
\hline
Parameterisation & $r_{E;p}$ (fm) & $r_{M;p}$ (fm) & $d_{\rm max}$ for $G^p_E$ (\%) & $d_{\rm max}$ for $G^p_M$ (\%) \\
\hline
\hline
`Standard dipole' & $0.8112$ & $0.8112$ & $0$ & $0$\\
A1 & $0.8780$ & $0.7681$ & $2.06$ & $2.58$\\
VAMZ & $0.8776$ & $0.8598$ & $1.36$ & $0.95$\\
YAHL & $0.8790$ & $0.8510$ & $2.30$ & $1.13$\\
\hline
\end{tabular}
\end{center}
\vspace{0.5cm}
\end{table}

Of course, given that a single-dipole model is used in the `standard dipole' parameterisations of $G^p_E$ and $G^p_M$, the perfect agreement between the results of this work and those obtained with the first parameterisation 
in Table \ref{tab:Comparison} is expected. As it does not assume that $r_{E;p} = r_{M;p}$, the parameterisation of this work is more general. The interest in Table \ref{tab:Comparison} lies in the comparisons of the 
results in the remaining three cases.

It appears that the agreement between the parameterisation of this work and the VAMZ parameterisation is close to the $1~\%$ level. Regarding the comparison with the YAHL parameterisation, the difference is larger in case of 
$G^p_E$, whereas it remains close to $1~\%$ in case of $G^p_M$. The maximal differences in case of the comparison with the A1 solution are between $2.0$ and $2.6~\%$, larger in case of $G^p_M$. To summarise, the differences 
between the solutions of Refs.~\cite{Bernauer2010,Venkat2011,Ye2018} and the results obtained with the parameterisation of this work (after the appropriate fixation of the model parameters $r_{E;p}$ and $r_{M;p}$) are 
comparable to the differences among the solutions of Refs.~\cite{Bernauer2010,Venkat2011,Ye2018} themselves in the $Q^2$ domain of this work.

In representative PSAs of the low-energy $\pi N$ data with the ETH model, the median relative uncertainty in the fitted values of the model parameters is about $5~\%$. In addition, the maximal relative uncertainties in the 
model predictions for the $\pi^+ p$ differential cross sections between $20$ and $45$ MeV range between $5.4$ and $6.1~\%$; the corresponding uncertainties in the model predictions for the $\pi^- p$ elastic-scattering 
differential cross sections are considerably larger, reaching $30~\%$ in backward scattering. In comparison, the aforementioned form-factor effects are small. The results of Table \ref{tab:Comparison} indicate that the PSAs 
of the low-energy $\pi N$ data with the ETH model can be made self-sufficient by replacing the imported parameterisations of the proton form factors by the simpler scheme of Section \ref{sec:ThisWork}. Provided that reliable 
$r_{E;p}$ and $r_{M;p}$ values are supplied, the parameterisation of this work suffices for the purposes of the PSAs of the low-energy $\pi N$ data with the ETH model. At present however, the fixation of $r_{E;p}$ and $r_{M;p}$ 
is not as straightforward as one might expect one century after the proton was given a name. I will next elaborate on this issue.

Before 2010, the estimates for the rms electric charge radius of the proton were predominantly based on the results of analyses of experimental data - (mostly) differential cross sections and (occasionally also) polarisation 
measurements - from $e p$ elastic scattering; those estimates hovered around $0.875$ fm. To the best of my knowledge, the first indications that something might be amiss about the values, recommended both by the PDG as well 
as by NIST (the former were mostly fixed from the CODATA compilations of the latter), appeared in a 2007 paper by Belushkin, Hammer, and Mei{\ss}ner \cite{Belushkin2007}. After using two theoretical approaches in a 
dispersion-relation analysis of the $e p$ experimental data, also employing the theoretical constraints of analyticity and unitarity, the authors reported two results ($0.830^{+0.005}_{-0.008}$ and $0.844^{+0.008}_{-0.004}$ 
fm), neither of which tallied well with the recommended (at that time) $r_{E;p}$ values, namely $0.8750(68)$ fm (PDG2006) and $0.8768(69)$ fm (CODATA2006).

A few years later, the pioneering experiment by Pohl and collaborators enabled the extraction of a precise $r_{E;p}$ value from muonic hydrogen \cite{Pohl2010}. Being heavier than electrons, muons come closer to the 
hydrogen nucleus, the consequence being that several small effects (e.g., effects pertaining to vacuum polarisation, to the fine/hypefine splitting, to the proton size, etc.) yield a larger (in comparison with the electronic 
hydrogen) difference in the binding energies of the $2S$ and $2P$ states (Lamb shift). The result of that experiment was: $r_{E;p} = 0.84184(36)(56)$ fm. A second, even more precise value from a follow-up experiment (also on 
muonic hydrogen) became available in 2013 \cite{Antognini2013}, confirming the earlier result: $r_{E;p} = 0.84087(26)(29)$ fm. Several turbulent years followed, during which attempts were made towards a resolution of what 
became known as `the proton-radius puzzle' on the basis of established or (more frequently) exotic Physics.

Between 2013 and 2019, the PDG retained the neutral (and somewhat awkward) stand of reporting both results (i.e., the CODATA2012/2016 results, as well as the results of Ref.~\cite{Antognini2013}) in their compilations, 
encouraging the experimentalists to settle the obvious discrepancy. As the 2016 results from muonic deuterium \cite{Pohl2016}, from electronic hydrogen~\footnote{A 2018 result \cite{Fleurbaey2018} is in favour of a high 
$r_{E;p}$ value, thus contradicting the results obtained from the other two experiments on electronic hydrogen.} \cite{Beyer2017,Bezginov2019} in 2017 and 2019, as well as from re-analyses of the $e p$ elastic-scattering data 
\cite{Lorenz2015,Griffioen2016,Higinbotham2016,Horbatsch2017,Alarcon2019} all pointed in the direction of a `low' $r_{E;p}$ value, the CODATA2016 result was dropped in the recent PDG compilation; recommended now by the PDG is 
$r_{E;p}=0.8409(4)$ fm \cite{pdg2020}, an average obtained from the results of Refs.~\cite{Antognini2013,Bezginov2019,Xiong2019}; these three results originate from measurements of the Lamb shift in muonic and electronic 
hydrogen, as well as from a recent $e p$ elastic-scattering experiment (`PRad' - Proton Radius Experiment at the Thomas Jefferson National Accelerator Facility) at low $Q^2$, respectively. The NIST also adapted their 
recommendation to $r_{E;p}=0.8414(19)$ fm (CODATA2018 \cite{CODATA2018}). All would have been perfect, had it not been for one hitch. In a 2019 paper, Hagelstein and Pascalutsa \cite{Hagelstein2019} demonstrated that a lower 
bound for $r_{E;p}$ can be obtained from the $e p$ elastic-scattering data; with $95~\%$ confidence, the lowest acceptable $r_{E;p}$ value appears to be equal to $0.847$ fm, i.e., a value exceeding both results from muonic 
hydrogen \cite{Pohl2010,Antognini2013} by several standard deviations.

There is no doubt that the incompatibility between the results of Refs.~\cite{Pohl2010,Antognini2013} and the lower bound obtained in Ref.~\cite{Hagelstein2019} calls for further investigation. In addition, it is imperative 
to understand the source of the discrepancy between the former results from the $e p$ elastic-scattering data and the currently-recommended values. These two comments have appeared in several other works.

Equally confusing is the available information on the rms magnetic radius of the proton $r_{M;p}$. The recommended value by the PDG between 2011 and 2015 was the 2010 result by the A1 Collaboration \cite{Bernauer2010}, whereas 
between 2016 and 2018 the PDG favoured a similar result obtained in Ref.~\cite{Lee2015} also from an analysis of the data from Mainz. In fact, two results had been obtained in Ref.~\cite{Lee2015}: $0.776(34)(17)$ fm from the 
data acquired in Mainz and $0.914(35)$ fm from the data acquired elsewhere~\footnote{Incidentally, `World data' as a description of `data acquired anywhere but Mainz' is misleading.}. Although these two results are incompatible 
(the p-value, corresponding to their reproduction by one constant, is equal to about $7.57 \cdot 10^{-3}$), Ref.~\cite{Lee2015} reported ``a simple average'' of $0.851(26)$ fm, which (surprisingly) the PDG adopted in their 2019 
compilation. However, when fitting a constant to incompatible measurements, it is imperative to correct the fitted uncertainties for the (poor) quality of the fit via the application of the Birge factor (which, in this case, 
comes out equal to about $2.67$); if not, the resulting uncertainty is not representative of the variation of the input values. The correct weighted average of the two incompatible $r_{M;p}$ results of Ref.~\cite{Lee2015} is 
not $0.851(26)$, but $0.851(69)$ fm!

\begin{figure}
\begin{center}
\includegraphics [width=15.5cm] {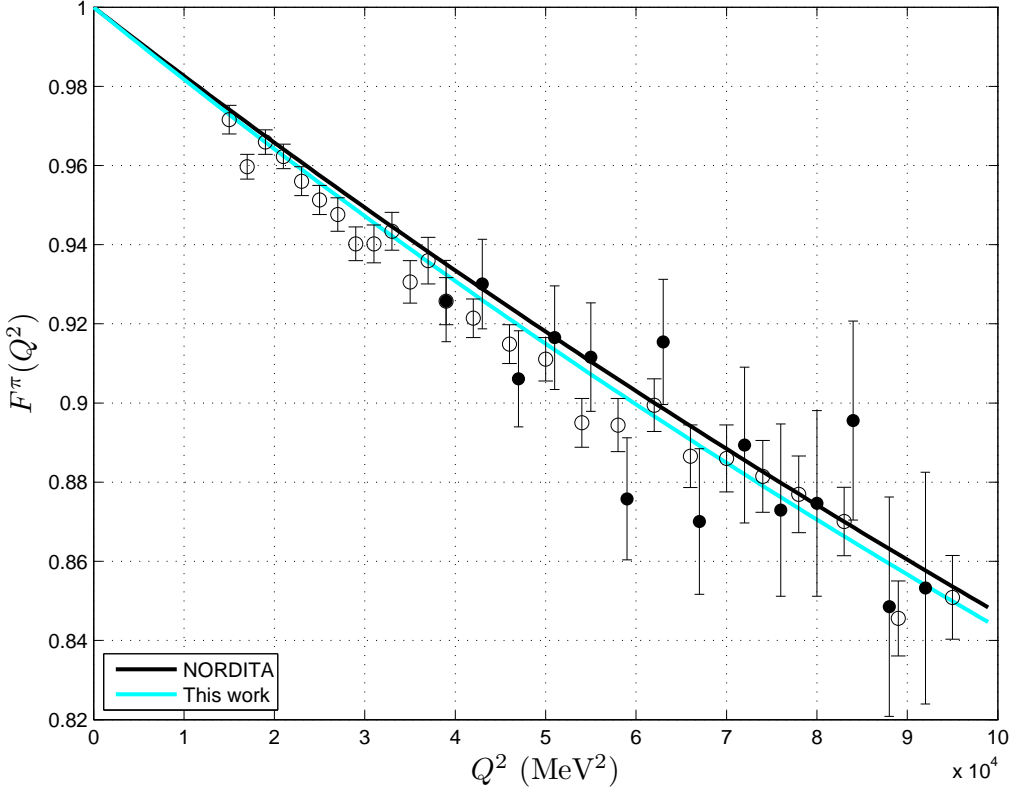}
\caption{\label{fig:Fpi}The pion form factor $F^\pi$. The NORDITA team defined \cite{Tromborg1974}: $F^\pi (t) = F^p_1 (t) - F^n_1 (t)$. The parameterisation of this work rests upon Eqs.~(\ref{eq:EQ0016},\ref{eq:EQ0017}) with 
$r_{E;\pi}=0.659$ fm \cite{pdg2020}. The data shown come from two experiments: the filled points have been taken from Ref.~\cite{Dally1982}, the open ones from Ref.~\cite{Amendolia1986}. In both cases, only the statistical 
uncertainties are shown.}
\vspace{0.35cm}
\end{center}
\end{figure}

For the sake of completeness, I will next give some results for the pion form factor $F^\pi$, which also enters the EM part of the $\pi N$ scattering amplitude of the ETH model. As Fig.~\ref{fig:Fpi} demonstrates, the agreement 
between the monopole model of this work (using $r_{E;\pi}=0.659$ fm \cite{pdg2020}) and the NORDITA parameterisation of $F^\pi$ is satisfactory. The data shown in the figure have been taken from Refs.~\cite{Dally1982,Amendolia1986}, 
which reported $\lvert F^\pi \rvert^2$ at forty (in total) $Q^2$ values below $0.1$ GeV$^2$.

The treatment of the normalisation effects in the experiment of Ref.~\cite{Dally1982} according to the Arndt-Roper method \cite{Arndt1972} (see Appendix \ref{App:AppA}) yields $r_{E;\pi} = 0.663(23)$ fm, which is the value 
reported in Ref.~\cite{Dally1982}. In an obvious attempt to demonstrate the near model-independence of their $r_{E;\pi}$ result, several estimates were extracted (and reported) in Ref.~\cite{Amendolia1986}, all accompanied 
by smaller uncertainties than those obtained in this work from the same set of data: from a monopole fit with constrained normalisation, the authors obtained $0.657(8)$ fm; from a monopole fit with free normalisation, they 
obtained $0.653(8)$ fm; from a dipole fit, they obtained $0.637(8)$ fm; finally, using a Pad{\'e}-type parameterisation (which they evidently favoured), they obtained $0.663(6)$ fm (this result was unquestionably imported 
into the PDG database \cite{pdg2020}). Although plurality can be desirable on several occasions, I find it confusing in this case. The use of the Arndt-Roper formula in the optimisation of the data of Ref.~\cite{Amendolia1986}, 
following the monopole approximation, yields the result: $r_{E;\pi} = 0.664(11)$ fm, end of story.

I subsequently pursued a common analysis of the two datasets \cite{Dally1982,Amendolia1986} and obtained the result: $r_{E;\pi} = 0.664(10)$ fm, as well as fitted scale factors which were close to $1$, namely equal to $0.9987$ 
and $0.9938$ for the datasets of Refs.~\cite{Dally1982,Amendolia1986}, respectively. Owing to the fact that the normalisation uncertainties in the two experiments were reported as $1.0~\%$ \cite{Dally1982} and $0.9~\%$ 
\cite{Amendolia1986}, the differences of two fitted scale factors to $1$ are well within the reported normalisation uncertainties.

\section{\label{sec:Conclusions}Conclusions}

The first goal in this work was the comparison of the results obtained from four parameterisations of the Sachs form factors of the proton $G^p_E$ (electric) and $G^p_M$ (magnetic) in the region of interest to a pion-nucleon 
($\pi N$) interaction model (ETH model) \cite{Goudsmit1994,Matsinos2006,Matsinos2014,Matsinos2017}, namely for $4$-momentum transfer $Q^2 \lesssim 0.1$ GeV$^2$. Compared were the results obtained
\begin{itemize}
\item from the `standard dipole' (Section \ref{sec:StandardDipole}), which had provided the routine parameterisation of the nucleon form factors up to the 1980s;
\item from a spline fit to electron-proton cross-section and polarisation measurements (Section \ref{sec:A1}); and
\item from two solutions from Arrington's teams (Sections \ref{sec:VAMZ} and \ref{sec:YAHL}).
\end{itemize}
The relative differences between these solutions remain smaller than about $2.8~\%$ in the aforementioned $Q^2$ region of interest.

The second goal herein was to introduce - and, to an extent, test - a parameterisation of the Sachs form factors of the proton $G^p_E$ and $G^p_M$ based on two dipoles, one pertaining to $G^p_E$ (admitting as parameter the rms 
electric charge radius of the proton $r_{E;p}$), the other to $G^p_M$ (admitting as parameter the rms magnetic radius of the proton $r_{M;p}$). Comparisons between (on the one hand) the four aforementioned parameterisation 
schemes of $G^p_E$ and $G^p_M$ at low $Q^2$ and (on the other hand) the parameterisation of this work were enabled after fixing the model parameters $r_{E;p}$ and $r_{M;p}$ to the appropriate values corresponding to each of the 
four parameterisations of Sections \ref{sec:StandardDipole}-\ref{sec:YAHL}. The resulting differences in the aforementioned $Q^2$ region of interest remained below about $2.6~\%$ (see Table \ref{tab:Comparison}), i.e., slightly 
below the maximal differences found when comparing the four parameterisations of Sections \ref{sec:StandardDipole}-\ref{sec:YAHL} among themselves.

The conclusion from these comparisons is that the parameterisation of this work suffices for the purposes of the ETH model. The replacement of the form factors, which the ETH model used after 2015, by the simple parameterisation 
of this work will remove from the phase-shift analyses (PSAs) of the low-energy $\pi N$ data with the ETH model the largest part of the dependence on extraneous sources.
 
The decision regarding the fixation of the two model parameters, $r_{E;p}$ and $r_{M;p}$, may be postponed to the time when the next PSA of the ETH model will be conducted.

\begin{ack}
I am indebted to M.~Horbatsch and to J.C.~Bernauer for their prompt response to my questions.

Figure \ref{fig:FeynmanGraphsETHZ} has been drawn with the software package JaxoDraw \cite{Binosi2004,Binosi2009}, available from jaxodraw.sourceforge.net. The remaining figures have been created with MATLAB$^{\textregistered}$~(The 
MathWorks, Inc., Natick, Massachusetts, United States).
\end{ack}

\clearpage
\newpage
\appendix
\section{\label{App:AppA}Formal treatment of datasets with known normalisation uncertainty}

The formal procedure for treating datasets, which are subject to normalisation uncertainty, rests upon the use of the Arndt-Roper formula \cite{Arndt1972}, see Refs.~\cite{Matsinos2006,Matsinos2017} (and several other 
references therein). According to this method, one parameter is introduced per dataset, to account for the fact that the absolute normalisation of each dataset is known with a finite (non-zero) uncertainty. This parameter, 
named normalisation parameter in Ref.~\cite{Arndt1972} and scale factor in Refs.~\cite{Matsinos2006,Matsinos2017}, is applied to each input dataset as a whole: all datapoints of a dataset are affected by the same (relative) 
amount.

The determination of the absolute normalisation of each dataset involves a `calibration' procedure, resting upon a comparison of experimental results of the reaction at issue with those of a reaction whose absolute 
normalisation is more accurately known. This comparison introduces one additional uncertainty, the normalisation uncertainty, which encompasses all known uncertainties associated with the calibration procedure. In practice, 
the fixation of the absolute normalisation of each dataset may be thought of as resting upon \emph{one} measurement, which is accompanied by an uncertainty, as all other datapoints of the dataset. Consequently, the treatment 
of the normalisation uncertainty in a manner which is different to that of the uncertainties of any of the datapoints of the dataset is hardly justifiable.

As I have found several statements in the literature, expressing discomfort at the `free' floating of the datasets and the introduction of `the additional parameters' it entails, I rather doubt that it is generally understood 
that the `floating' of the datasets is not `free', but `controlled', in that it is accompanied by an appropriate compensation to the minimisation function. The $\chi^2$ contributions of each dataset arise from the differences 
between the rescaled fitted values and the input values, as well as from a term taking account of the departure of the scale factor of the dataset from $1$. As each scale factor appears only in the $\chi^2$ contribution 
of one dataset, the minimisation of the overall $\chi^2$ (with respect to each scale factor) is equivalent to the minimisation of the $\chi^2$ contribution of each dataset (with respect to its own scale factor). This requirement 
\emph{fixes} each scale factor from the fitted and the input values at \emph{each} step of the optimisation.

To conclude, it is true that one additional parameter per dataset is introduced in the optimisation when the floating of the datasets is allowed. However, this parameter is fixed at each step of the optimisation. Consequently, 
one ends up with exactly the same number of degrees of freedom in the problem as when the normalisation effects are altogether ignored (i.e., when no floating of the input datasets is allowed). The interested reader is 
referred to Refs.~\cite{Matsinos2006,Matsinos2017}.


\begin{thebibliography}{99}

\bibitem{Aitchison2013} I.J.R.~Aitchison, A.J.G.~Hey, `Gauge Theories in Particle Physics: A Practical Introduction, Volume 1: From Relativistic Quantum Mechanics to QED', 4th Edn, Taylor \& Francis Group, LLC (2013). ISBN-13: 978-1-4665-1302-0
\bibitem{pdg2020} P.A.~Zyla \etal~(Particle Data Group), `The Review of Particle Physics (2020)', Prog.~Theor.~Exp.~Phys.~2020, 083C01 (2020).
\bibitem{Schneider17} G.~Schneider \etal, `Double-trap measurement of the proton magnetic moment at $0.3$ parts per billion precision', Science 358, 1081 (2017). DOI: 10.1126/science.aan0207
\bibitem{Rosenbluth1950} M.N.~Rosenbluth, `High energy elastic scattering of electrons on protons', Phys.~Rev.~79, 615 (1950). DOI: 10.1103/PhysRev.79.615
\bibitem{Hand1963} L.N.~Hand, D.G.~Miller, R.~Wilson, `Electric and magnetic form factors of the nucleon', Rev.~Mod.~Phys.~35, 335 (1963). DOI: 10.1103/RevModPhys.35.335
\bibitem{Mohr2016} P.J.~Mohr, D.B.~Newell, B.N.~Taylor, `CODATA Recommended Values of the Fundamental Physical Constants: 2014', J.~Phys.~Chem.~Ref.~Data 45, 043102 (2016). DOI: 10.1063/1.4954402
\bibitem{Goudsmit1994} P.F.A.~Goudsmit, H.J.~Leisi, E.~Matsinos, B.L.~Birbrair, A.B.~Gridnev, `The extended tree-level model of the pion-nucleon interaction', Nucl.~Phys.~A 575, 673 (1994). DOI: 10.1016/0375-9474(94)90162-7
\bibitem{Matsinos2006} E.~Matsinos, W.S.~Woolcock, G.C.~Oades, G.~Rasche, A.~Gashi, `Phase-shift analysis of low-energy $\pi^\pm p$ elastic-sacttering data', Nucl.~Phys.~A 778, 95 (2006). DOI: 10.1016/j.nuclphysa.2006.07.040
\bibitem{Matsinos2014} E.~Matsinos, G.~Rasche, `Aspects of the ETH model of the pion-nucleon interaction', Nucl.~Phys.~A 927, 147 (2014).\\DOI: 10.1016/j.nuclphysa.2014.04.021
\bibitem{Matsinos2017} E.~Matsinos, G.~Rasche, `Update of the phase-shift analysis of the low-energy $\pi N$ data', {\tt arXiv:1706.05524 [nucl-th]}.
\bibitem{Matsinos2020a} E.~Matsinos, `Determination of the masses and decay widths of the scalar-isoscalar and vector-isovector mesons below $2$ GeV', {\tt arXiv:2007.13130 [hep-ph]}.
\bibitem{Matsinos2020b} E.~Matsinos, `Determination of the masses and decay widths of the well-established $s$ and $p$ baryon resonances below $2$ GeV', {\tt arXiv:2008.06919 [hep-ph]}.
\bibitem{Hill2010} R.J.~Hill, G.~Paz, `Model-independent extraction of the proton charge radius from electron scattering', Phys.~Rev.~D 82, 113005 (2010). DOI: 10.1103/PhysRevD.82.113005
\bibitem{Sick2018} I.~Sick, `Proton charge radius from electron scattering', Atoms 6(1), 2 (2018). DOI: 10.3390/atoms6010002
\bibitem{Sick2014} I.~Sick, D.~Trautmann, `Proton root-mean-square radii and electron scattering', Phys.~Rev.~C 89, 012201(R) (2014). DOI: 10.1103/PhysRevC.89.012201
\bibitem{Lachniet2009} J.~Lachniet \etal~(CLAS Collaboration), `A precise measurement of the neutron magnetic form factor $G^n_M$ in the few-GeV$^2$ region', Phys.~Rev.~Lett.~102, 192001 (2009). DOI: 10.1103/PhysRevLett.102.192001
\bibitem{Tromborg1976} B.~Tromborg, S.~Waldestr{\o}m, I.~{\O}verb{\o}, `Electromagnetic corrections to $\pi^+ p$ scattering', Ann.~Phys.~100, 1 (1976). DOI: 10.1016/0003-4916(76)90055-5
\bibitem{Tromborg1977} B.~Tromborg, S.~Waldestr{\o}m, I.~{\O}verb{\o}, `Electromagnetic corrections to $\pi N$ scattering', Phys.~Rev.~D 15, 725 (1977). DOI: 10.1103/PhysRevD.15.725
\bibitem{Tromborg1978} B.~Tromborg, S.~Waldestr{\o}m, I.~{\O}verb{\o}, `Electromagnetic corrections in hadron scattering, with application to $\pi N \to \pi N$', Helv.~Phys.~Acta 51, 584 (1978).
\bibitem{Tromborg1974} B.~Tromborg, J.~Hamilton, `Electromagnetic corrections to hadron-hadron scattering', Nucl.~Phys.~B 76, 483 (1974). DOI: 10.1016/0550-3213(74)90538-0
\bibitem{Bernauer2010} J.C.~Bernauer \etal~(A1 Collaboration), `High-precision determination of the electric and magnetic form factors of the proton', Phys.~Rev.~Lett.~105, 242001 (2010). DOI: 10.1103/PhysRevLett.105.242001
\bibitem{Bernauer2014} J.C.~Bernauer \etal~(A1 Collaboration), `Electric and magnetic form factors of the proton', Phys.~Rev.~C 90, 015206 (2014). DOI: 10.1103/PhysRevC.90.015206
\bibitem{Arrington2007} J.~Arrington, W.~Melnitchouk, J.A.~Tjon, `Global analysis of proton elastic form factor data with two-photon exchange corrections', Phys.~Rev.~C 76, 035205 (2007). DOI: 10.1103/PhysRevC.76.035205
\bibitem{Blunden2003} P.G.~Blunden, W.~Melnitchouk, J.A.~Tjon, `Two-photon exchange and elastic electron-proton scattering', Phys.~Rev.~Lett.~91, 142304 (2003). DOI: 10.1103/PhysRevLett.91.142304
\bibitem{Venkat2011} S.~Venkat, J.~Arrington, G.A.~Miller, X.~Zhan, `Realistic transverse images of the proton charge and magnetization densities', Phys.~Rev.~C 83, 015203 (2011). DOI: 10.1103/PhysRevC.83.015203
\bibitem{Ye2018} Z.~Ye, J.~Arrington, R.J.~Hill, G.~Lee, `Proton and neutron electromagnetic form factors and uncertainties', Phys.~Lett.~B 777, 8 (2018). DOI: 10.1016/j.physletb.2017.11.023
\bibitem{Belushkin2007} M.A.~Belushkin, H.-W.~Hammer, Ulf-G.~Mei{\ss}ner, `Dispersion analysis of the nucleon form factors including meson continua', Phys.~Rev.~C 75, 035202 (2007). DOI: 10.1103/PhysRevC.75.035202
\bibitem{Pohl2010} R.~Pohl \etal, `The size of the proton', Nature 466, 213 (2010). DOI: 10.1038/nature09250
\bibitem{Antognini2013} A.~Antognini \etal, `Proton structure from the measurement of $2S - 2P$ transition frequencies of muonic hydrogen, Science 339, 417 (2013). DOI: 10.1126/science.1230016
\bibitem{Pohl2016} R.~Pohl \etal, `Laser spectroscopy of muonic deuterium', Science 353, 669 (2016). DOI: 10.1126/science.aaf2468
\bibitem{Fleurbaey2018} H.~Fleurbaey \etal, `New measurement of the $1S - 3S$ transition frequency of hydrogen: Contribution to the proton charge radius puzzle', Phys.~Rev.~Lett.~120, 183001 (2018).\\DOI: 10.1103/PhysRevLett.120.18300110.1103/PhysRevLett.120.183001
\bibitem{Beyer2017} A.~Beyer \etal, `The Rydberg constant and proton size from atomic hydrogen', Science 358, 79 (2017). DOI: 10.1126/science.aah6677
\bibitem{Bezginov2019} N.~Bezginov \etal, `A measurement of the atomic hydrogen Lamb shift and the proton charge radius', Science 365, 1007 (2019). DOI: 10.1126/science.aau7807
\bibitem{Lorenz2015} I.T.~Lorenz, Ulf-G.~Mei{\ss}ner, H.-W.~Hammer, Y.-B.~Dong, `Theoretical constraints and systematic effects in the determination of the proton form factors', Phys.~Rev.~D 91, 014023 (2015). DOI: 10.1103/PhysRevD.91.014023
\bibitem{Griffioen2016} K.~Griffioen, C.~Carlson, S.~Maddox, `Consistency of electron scattering data with a small proton radius', Phys.~Rev.~C 93, 065207 (2016). DOI: 10.1103/PhysRevC.93.065207
\bibitem{Higinbotham2016} D.W.~Higinbotham \etal, `Proton radius from electron scattering data', Phys.~Rev.~C 93, 055207 (2016). DOI: 10.1103/PhysRevC.93.055207
\bibitem{Horbatsch2017} M.~Horbatsch, E.A.~Hessels, A.~Pineda, `Proton radius from electron-proton scattering and chiral perturbation theory', Phys.~Rev.~C 95, 035203 (2017). DOI: 10.1103/PhysRevC.95.035203
\bibitem{Alarcon2019} J.M.~Alarc{\'o}n, D.W.~Higinbotham, C.~Weiss, Z.~Ye, `Proton charge radius extraction from electron scattering data using dispersively improved chiral effective field theory', Phys.~Rev.~C 99, 044303 (2019). DOI: 10.1103/PhysRevC.99.044303
\bibitem{Xiong2019} W.~Xiong \etal~(PRad Collaboration), `A small proton charge radius from an electron-proton scattering experiment', Nature 575, 147 (2019). DOI: 10.1038/s41586-019-1721-2
\bibitem{CODATA2018} https://physics.nist.gov/cgi-bin/cuu/Value?rp, accessed on September 5, 2020.
\bibitem{Hagelstein2019} F.~Hagelstein, V.~Pascalutsa, `Lower bound on the proton charge radius from electron scattering data', Phys.~Lett.~B 797, 134825 (2019). DOI: 10.1016/j.physletb.2019.134825
\bibitem{Lee2015} G.~Lee, J.R.~Arrington, R.J.~Hill, `Extraction of the proton radius from electron-proton scattering data', Phys.~Rev.~D 92, 013013 (2015). DOI: 10.1103/PhysRevD.92.013013
\bibitem{Dally1982} E.B.~Dally \etal, `Elastic-scattering measurement of the negative-pion radius', Phys.~Rev.~Lett.~48, 375 (1982). DOI: 10.1103/PhysRevLett.48.375
\bibitem{Amendolia1986} S.R.~Amendolia \etal~(NA7 collaboration), `A measurement of the space-like pion electromagnetic form factor', Nucl.~Phys.~B 277, 168 (1986). DOI: 10.1016/0550-3213(86)90437-2
\bibitem{Arndt1972} R.A.~Arndt, L.D.~Roper, `The use of partial-wave representations in the planning of scattering measurements. Application to $330$ MeV $n p$ scattering', Nucl.~Phys.~B 50 (1972) 285--300. DOI: 10.1016/S0550-3213(72)80019-1
\bibitem{Binosi2004} D.~Binosi, L.~Theu\ss{}l, `JaxoDraw: A graphical user interface for drawing Feynman diagrams', Comput.~Phys.~Commun.~161 (2004) 76--86. DOI: 10.1016/j.cpc.2004.05.001
\bibitem{Binosi2009} D.~Binosi, J.~Collins, C.~Kaufhold, L.~Theu\ss{}l, `JaxoDraw: A graphical user interface for drawing Feynman diagrams. Version 2.0 release notes', Comput.~Phys.~Commun.~180 (2009) 1709--1715.\\DOI: doi.org/10.1016/j.cpc.2009.02.020

\end{thebibliography}
\end{document}